\documentclass[11pt,twoside]{article}

\usepackage{asp2006}
\usepackage{epsf}
\usepackage{psfig}
\usepackage{lscape}

\begin{document}
\title{Detection of Endolithes Using Infrared Spectroscopy}
\author{S. Dumas, Y. Dutil and G. Joncas}   %%% Fill in author names
\affil{Dept. de physique, de g\'enie physique et d'optique et Observatoire du mont M\'egantic, Univsersit\'e Laval, Qu\'ebec, Canada, G1K 7P4}  %%% Fill in author affiliations

\begin{abstract} %%% Abstract to run on from here.

On Earth, the Dry Valleys of Antarctica provide the closest martian-like environment for the study of extremophiles. Colonies of bacteries are protected from the freezing temperatures, the drought and UV light. They represent almost half of the biomass of those regions. Due to there resilience, endolithes are one possible model of martian biota. 

We propose to use infrared spectroscopy to remotely detect those colonies even if there is no obvious sign of their presence. This remote sensing approach reduces the risk of contamination or damage to the samples. 

\end{abstract}

\section{Introduction}

Space exploration is a difficult task and the search for life is no different. The equipement size a probe can bring severely limit the scope of the search. Every sample cannot be analyzed and selecting those that can be is not trivial. This project investigates the possibilities of remote detection using infrared spectroscopy in order to select those few samples to pick and analyse further.

\section{Infrared Spectroscopy}

The reason for using IR spectroscopy is mainly that almost all biological marker molecules (i.e. biomarkers) will show some spectral features in the near to far IR (\cite{hand2005}). Furthermore, the spectroscopy will not destroy the sample (e.g. there is no contact with the sample, no contamination). Previous techniques of detection tended to destroy, or damage, the endolithes or the environment in which they lived (e.g. by using electron-microscope, chemical and biological or analysis).

Biomarkers are an important source of information in the search for evidence of life in geological samples. Even if the organisms are dead or dormant, it is still possible to detect their presence, byproducts or even their chemical alterations of the environment.

In searching for extraterrestrial life, it is important to have a minimum of preconception about them in order to find it.

\section{Endolithes}

Endolithes are organisms that live inside rocks or in the pores between mineral grains. There are thousands of known species of endolithes, including members from Bacteria, Archaea, and Fungi.  They represent near half the Earth's biomass and also present an ideal model of life for Mars (\cite{ascaco2002}, \cite{hand2005}).

This study used two groups of samples. One from the Guelph region (west of Toronto, Ontario) and the other near Eureka on the Ellesmere Island in the Nunavut (\cite{omelon2006}). Both regions have different geology and climate.

\section{Methodology}

Samples were scanned using a technique called diffuse scattering. The IR beam was directed toward the sample and a series of mirrors redirected the scattered beam to the IR sensor. The spectra were obtained using an IR Nicolet spectrometer.

The apparatus used to collect the scattered IR beam could not receive large rock samples. It was necessary to break them into smaller pieces. The same device was very sensible to the angle of incidence and reflectance. It was important to position both mirrors near the vertical above the sample. The adjustment of each mirrors, in order to optimize the reception of light, was very time consuming. Those adjustments were performed using an aluminum plate as the sample to maximize the received flux to the sensor. When the rock sample was placed in the light path, the total flux drops a lot but the signal-to-noise ratio was high enough.

Spectra in middle IR (132) and near IR (49) on 15 samples from the two groups were taken. The best results were from the Middle IR. Most of the spectra were taken at a resolution of 4 or 8 cm$^{-1}$. After analysis, it appears that a spectral resolution of 8 cm$^{-1}$ is enough for our purpose. The middle IR spectra (from 4000 to 650 cm$^{-1}$) were then processed using Principal Component Analysis in order to classify them.

\section{Principal Component Analysis}

Principal Component Analysis (\cite{marchi2007}) is a technique of Factorial Analysis (e.g. Multivariate Statistics). It is often used to find a new coordinate system in which the original data will be better aligned on some axes (e.g. principal components). 

The technique can be summarized by the equation $A=UWV^T$ where A,U,W and V are matrices. 

{\em A} is a $m \times n$ matrix containing the spectra on each row. The value of each absorption band is then regrouped in the columns which are the variables on which the PCA works. {\em U}, also a $m \times n$ matrix, contains the spectra in the new coordinate system. The matrix {\em V}, where $VV^{T}$ is an identity matrix, contains the eigenvectors. {\em W} is a diagonal matrix containing the squared root of the eigenvalues and provides a clue about how many principal components (PC) can be used to describe the data. It is important to have a matrix {\em A} with more columns than rows (ie. $n>m$) else {\em W} becomes a singular matrix and the whole PCA fails.

To facilitate data manipulation and visualization for this paper, we have taken only the first three principal components. The result of the PCA process is illustrated in figure 1. The process can be extended to as many components as needed. Based on the eigenvalues from our data, the first five components are relevant, the others being buried in the noise.

We have identified several clusters. Cluster {\em E} groups the spectra containing features showing the presence of organic compounds. Cluster {\em E} is very close to another cluster of points calculated from organic spectra used as a reference (i.e. cluster O). Further, most of the spectra of cluster {\em E} have been taken directly from green regions visible on the samples.

\begin{figure}
\begin{center}
\includegraphics{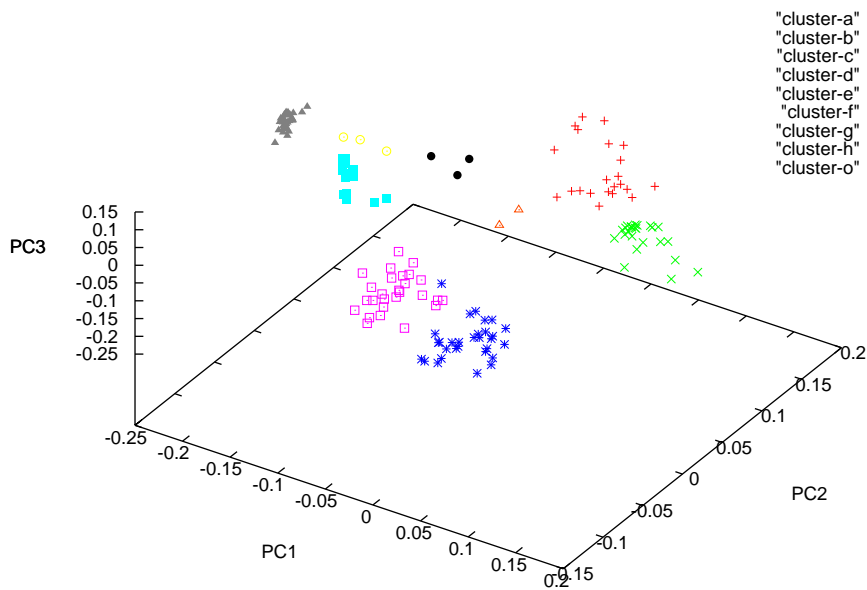} 
\end{center}
\caption{\textsl{Spectra plotted in the PCA space}}
\end{figure}

\section{Results and Conclusions}

Our results show that it is possible to detect biological signatures using a spectrometer operating in the middle infrared range. However, it is not possible to highlight a particular region, or regions, in a spectrum to be used to identify biomarkers. The interdependance of the absorption bands related to the living are too complex to simply isolate a few.

The proposed technique calls for a more subtle approach by comparing witness spectra and an unknown spectrum by plotting them in the PCA's space. If the test spectrum, once projected into the PCA's space, is close to the reference group then the probability of it is containing biomarkers is high. Adding a non-organic spectra to the PCA space as references may improve the idenfitication scheme.

This technique could be used to pinpoint potential life harboring rock for more detailed analysis. It could be possible to extend the method to better idenfity the unknown spectrum using a more precise reference database.

\acknowledgements %%% Text of acknowledgments runs on after this command.

We wish to thank the following people for their help in providing the samples used for this study : 
\begin{itemize}
 \item Christopher Omelon, Microbial Geochemistry Laboratory, University of Toronto Geology
 \item Uta Matthes, Department of Integrative Biology, University of Guelph
\end{itemize}

\end{document}